\documentclass[conference]{IEEEtran}
\IEEEoverridecommandlockouts

\usepackage[cmyk]{xcolor}
\usepackage[hidelinks]{hyperref}
\usepackage[utf8]{inputenc}
\usepackage{newtxtext,newtxmath}
\usepackage[american]{babel}
\usepackage{csquotes}
\usepackage[final]{microtype}
\usepackage{multirow}
\usepackage{xurl}
\usepackage{booktabs}
\usepackage{tabularx,ragged2e}
\newcolumntype{L}{>{\raggedright\arraybackslash}X}
\newcolumntype{R}{>{\raggedleft\arraybackslash}X}
\usepackage{threeparttable}
\usepackage{stmaryrd}
\usepackage{float}
\usepackage{datetime}

\usepackage{pgfplots}
\pgfplotsset{compat=1.17}
\usepackage[outline]{contour}
\contourlength{0.2em}
\usepackage{tikz}
\usetikzlibrary{patterns}
\usepgfplotslibrary{fillbetween}
\usetikzlibrary{patterns}
\usetikzlibrary{positioning, calc}

\usepackage{siunitx}
\DeclareSIUnit\tx{tx}
\DeclareSIPrefix{\noop}{}{0} % https://tex.stackexchange.com/a/605736/47127
\usepackage[backend=biber, style=ieee]{biblatex}
\usepackage{acro}
\DeclareAcronym{PoW}{
  short = PoW,
  long  = proof-of-work
}
\DeclareAcronym{PoS}{
  short = PoS,
  long  = proof-of-stake
}
\DeclareAcronym{DLT}{
  short = DLT,
  long  = distributed ledger technology
}
\DeclareAcronym{TPS}{
  short = tps,
  long  = transactions per second
}
\DeclareAcronym{GHG}{
  short = GHG,
  long  = greenhouse gas
}
\DeclareAcronym{ASIC}{
  short = ASIC,
  long  = application-specific integrated circuit
}

\DeclareAcronym{PPA}{
  short = PPA,
  long  = power purchase agreement
}
\DeclareAcronym{L1}{
  short = L1,
  long  = layer-1
}
\DeclareAcronym{L2}{
  short = L2,
  long  = layer-2
}
\DeclareAcronym{IEA}{
  short = IEA,
  long  = International Energy Agency
}
\DeclareAcronym{GWP}{
  short = GWP,
  long  = global warming potential
}
\DeclareAcronym{BTM}{
  short = BTM,
  long  = behind-the-meter
}
\DeclareAcronym{FTM}{
  short = FTM,
  long  = front-of-the-meter
}
\DeclareAcronym{ROI}{
  short = ROI,
  long  = return on investment
}
\DeclareAcronym{CLR}{
  short = CLR,
  long  = controllable load resource
}
\DeclareAcronym{NCLR}{
  short = NCLR,
  long  = non-controllable load resource
}
\DeclareAcronym{RECs}{
  short = RECs,
  long  = Renewable energy certificates
}
\DeclareAcronym{GOs}{
  short = GOs,
  long  = guarantees of origin
}
\DeclareAcronym{ETFs}{
  short = ETFs,
  long  = exchange-traded funds
}
\DeclareAcronym{CO2}{
  short = CO2,
  long  = carbon dioxide
}
\DeclareAcronym{US}{
  short = US,
  long  = United States
}
\DeclareAcronym{EU}{
  short = EU,
  long  = European Union
}
\DeclareAcronym{VRE}{
  short = VRE,
  long  = variable renewable energy
}
\DeclareAcronym{BTC}{
  short = BTC,
  long  = Bitcoin Core
}
\DeclareAcronym{BMC}{
  short = BMC,
  long  = Bitcoin Mining Council
}
\DeclareAcronym{BECI}{
  short = BECI,
  long  = Bitcoin Energy Consumption Index
}
\DeclareAcronym{RES}{
  short = RES,
  long  = renewable energy sources
}
\DeclareAcronym{RE}{
  short = RE,
  long  = renewable energy
}
\DeclareAcronym{VOCs}{
  short = VOCs,
  long  = volatile organic compounds
}
\DeclareAcronym{IPCC}{
  short = IPCC,
  long  = Intergovernmental Panel on Climate Change
}
\DeclareAcronym{OSTP}{
  short = OSTP,
  long  = Office of Science and Technology Policy
}
\DeclareAcronym{LCOE}{
  short = LCOE,
  long  = Levelized Cost of Energy
}
\DeclareAcronym{DR}{
  short = DR,
  long  = demand-response
}
\DeclareAcronym{FLR}{
  short = FLR,
  long  = flexible load response
}
\DeclareAcronym{PoC}{
  short = PoC,
  long  = Proof of Concept
}
\DeclareAcronym{EACs}{
  short = EACs,
  long  = energy attribution certificates
}
\DeclareAcronym{ESG}{
  short = ESG,
  long  = {environmental, social and governance}
}

\usepackage[nomessages]{fp}

\newif\ifdraft
\drafttrue
\ifdraft
\newcommand{\jinote}[1]{ {\textcolor{purple} { ***Juan: #1 }}}
\newcommand{\afnote}[1]{ {\textcolor{green} { ***Alex: #1 }}}

\else
\newcommand{\jinote}[1]{}
\newcommand{\afnote}[1]{}
\fi

\DeclareRobustCommand*{\IEEEauthorrefmark}[1]{%
  \raisebox{0pt}[0pt][0pt]{\textsuperscript{\footnotesize #1}}%
}

\begin{document}

\title{Don't Trust, Verify: Towards a Framework for the Greening of Bitcoin}

\author{
    \IEEEauthorblockN{
                Juan Ignacio Ibañez\IEEEauthorrefmark{1}\IEEEauthorrefmark{2}\IEEEauthorrefmark{3},
                Alexander Freier\IEEEauthorrefmark{2}\IEEEauthorrefmark{3}\IEEEauthorrefmark{4}
    }

    \IEEEauthorblockA{\IEEEauthorrefmark{1}DLT Science Foundation, London, UK}
    \IEEEauthorblockA{\IEEEauthorrefmark{2}Centre for Blockchain Technologies, University College London, London, UK}
    \IEEEauthorblockA{\IEEEauthorrefmark{3}Facultad de Ciencia Política y Relaciones Internacionales, Universidad Católica de Córdoba, Córdoba, Argentina}
    \IEEEauthorblockA{\IEEEauthorrefmark{4}Energiequelle GmbH, Feldheim, Germany}
    \IEEEauthorblockA{j.ibanez@ucl.ac.uk, freier@energiequelle.de
}
}

\maketitle

Version: \today

% 150 Word Limit!
\begin{abstract}
% Introduction. In one sentence, what’s the topic?
For more than a decade, Bitcoin has gained as much adoption as it has received criticism. 
% State the problem you tackle
Fundamentally, Bitcoin is under fire for the high carbon footprint that results from the energy-intensive \ac{PoW} consensus algorithm. There is a trend however for Bitcoin mining to adopt a trajectory toward achieving carbon-negative status, notably due to the adoption of methane-based mining and mining-based \ac{FLR} to complement \ac{VRE} generation. Miners and electricity sellers may increase their profitability not only by taking advantage of excess energy, but also by selling green tokens to buyers interested in greening their portfolios.
% Summarize (in one sentence) why nobody else has adequately answered the research question yet.
Nevertheless, a proper ``green Bitcoin" accounting system requires a standard framework for the accreditation of sustainable bitcoin holdings. The proper way to build such a framework remains contested.
% Explain, in one sentence, how you tackled the research question.
In this paper, we survey the different sustainable Bitcoin accounting systems.
% Results
Analyzing the various alternatives, we suggest a path forward.
\end{abstract}

% 5 Keyword Limit!
% Avoid keywords that appear in the title to increase the probability of being found

% Introduce problem, outline solution; the statement of the problem should include a clear statement why the problem is important (or interesting).
\section{Introduction}
\label{sec:introduction}

Spearheaded by the revolutionary Bitcoin protocol and its cryptocurrency bitcoin, blockchain technology is evolving beyond mere promise and becoming a reality. However, blockchain in general and Bitcoin in particular have been starkly criticized for the high-energy consumption of the \ac{PoW} consensus algorithm \cite{Platt2021TheProof-of-Work,Ibanez2023TheExpansion}. This is mostly due to concerns that mining could aggravate climate change. Nevertheless, a growing body of literature is beginning to highlight that Bitcoin has the potential to be not only decarbonized, but furthermore to act as a net-negative carbon asset \cite{Ibanez2023CanSoK,Freier2023BitcoinGermany}.

In short, the contention is that the process of Bitcoin mining has unique characteristics that make it an exceptionally good complement to renewable energy generation, as it provides a load that is flexible, interruptible, non-rival, available, stable and reliable, highly price sensitive, scale agnostic, and portable, while at the same time producing an output whose price is not correlated to electricity prices and provides an additional source of revenue for the renewable energy seller \cite{Ibanez2023CanSoK}. This complementarity is expected to grow over time as mining profits fall (due to increased competition in the mining market and the Bitcoin halving) and as renewable energy imbalances grow (due to renewable penetration) \cite{Ibanez2023CanSoK}.

Although the argument is plausible, whether such a phenomenon will act as a mere mitigation factor on Bitcoin's carbon footprint or as a true net-decarbonizing force for the energy grid as a whole remains an empirical issue. However, this does not mean that it is merely exogenous, as miners may have the ability to influence the protocol towards becoming the latter.

Specifically, if miners manage to monetize the green attributes associated with their activities, this may give them a competitive edge over other miners \cite{Freier2023BitcoinGermany}. In turn, this ¿ could act as a tipping point factor, tilting the scale towards decarbonization.

\subsection{State of the art and our contribution}
% Previous or obvious approach

Over time, a number of works have been produced seeking to give green Bitcoin miners an additional source of profit. Cross and Bailey \cite{Cross2021GreeningOffsets} systematized the effects of holding and mining bitcoin for sustainability, and suggested an incentive offset scheme on this basis. Martinez \cite{Martinez2023CleanWhitepaper} outlined a framework for carbon markets in Bitcoin that are consistent with existing \ac{ESG} reporting standards, whereas SBP \cite{SBP2023SustainableWhitepaper} did so by exploring the interplay of Bitcoin with \ac{EACs}. % Approach/solution/contribution
Nevertheless, no significant analysis systematizes the state of the art. We provide an up-to-date framework that fills this gap.

% Overview
To do so, this paper is structured in the following form. First, we review forms of Bitcoin carbon accounting, comparing different estimation methods. Second, we survey proposed schemes to account for and incentivize Bitcoin greenness, exploring their advantages and disadvantages. We then discuss the findings and extract the key conclusions.

\section{Carbon accounting with Bitcoin}
\label{sec:accounting}

The elaboration of a framework to account for green bitcoin attributes in terms of \ac{GHG} requires an understanding of carbon accounting for bitcoin emissions in the first place. We identify two sets of systems addressing different aspects of Bitcoin emissions. Firstly, systems that ascertain what percentage of the total emissions of a system are attributable to a particular subsystem (in this case the, Bitcoin protocol). Secondly, systems discussing the event that is causally associated with the emissions and which should act as the denominator for the emissions (emissions per event).

\subsection{Estimation methods for total emissions}

Calculating the total emissions of the Bitcoin protocol in a given grid requires the previous adoption of a philosophical position. Although in some cases of off-grid mining or arrangements between a miner and an energy seller that is putting in place additional generation to meet its obligations (giving rise in turn to external verification issues such as those associated with self-reporting), emissions accounting is not usually straightforward as it requires the prioritization of some emissions over others.

To analyze each method, we resort to the 3 carbon accounting constraints identified by Cross \cite{Cross2023NoConstraints}:
\begin{enumerate}
    \item "Compositionality. The sum of carbon footprints of actors in a system must equal the total footprint of the system.
    \item "Marginality. The footprint of an action must reflect the difference between that action being taken and (counterfactually) not being taken."
    \item "Fungibility. The footprint of any two users of the same energy over the same duration must be equal."
\end{enumerate}

\paragraph{Marginal emissions accounting}

The marginal emissions method is a form of a before-and-after causal inference. If a miner plugged into the grid starts their operations, the difference between the grid emissions after the start of their operations and the emissions before the start is attributable to the miner. This can also be formulated as a hypothetical ex-ante: the additional new emissions that the grid \textit{would have}, if the miner started operations, are attributable to the miner. This can also be performed in reverse: the difference between the emissions that the grid would have if the miner turned off and the emissions that the grid actually has are the emissions that should be imputed to the miner. This approximation tends to highlight the fact that the baseline energy needs are met with renewable energy and only the net load once renewable energy is exhausted is met with fossil energy. In other words, fossil energy sources are the marginal seller.

This approach is intuitive and is usually employed by Bitcoin critics, presumably because Bitcoin is a rather new source of energy demand \cite{Dance2023TheBitcoin}. However, there are four problems with it.
\begin{enumerate}
    \item It arbitrarily assigns "green energy" to the old energy consumers and "brown energy" to the new energy consumers, despite they are all consuming fungible electrons from the same energy grid, for no other reason than a first-come-first-served principle. Therefore, it sacrifices the fungibility of load (see discussion below).
    \item If applied consistently to all energy buyers, it leads to results where the total emissions of the grid are much smaller than the sum of the individual emissions of all buyers. Therefore, it sacrifices the compositionality of the load.
    \item It displays a very significant short-term bias that is inconsistent with how other loads are usually viewed, e.g. batteries. An additional load may create an incentive for renewable buildout (demand leads to investment) resulting in a new equilibrium not considered by the marginal emissions figure. For instance, a battery may have high marginal emissions while at the same time enabling the deployment of additional renewable capacity.
    \item Short-run marginal emissions may not coincide with long-run marginal emissions (and increasing marginal emissions may coexist with decreasing average emissions \cite{Holland2022WhyPolicy}). %See https://twitter.com/nlmorley/status/1641616217772544001
\end{enumerate}

\paragraph{Attributional accounting}

Attributional accounting is the form of carbon accounting that takes the totality of a grid's emissions and attributes them to all the energy consumers. By starting from the total grid emissions, it preserves compositionality. The simplest form of attributional accounting is a simple average (``average emissions"), which results in a form of grid mix egalitarianism. Ryan et al \cite{Ryan2016ComparativeEmissions} have identified numerous variants of attributional accounting. 

Average emission factors preserve fungibility as well. Overall, their usefulness relative to marginal emission factors depends on the research questions (for instance, average emission factors are considered more appropriate for the attribution of responsibility \cite{Ryan2016ComparativeEmissions}. However, they do not necessarily account for marginality, which is a useful notion.

\paragraph{Emissions based on \ac{EACs}}

Using \ac{EACs} such as \ac{RECs} or \ac{GOs} could be considered a form of attributional accounting.  However, due to their relevance to understanding the functioning of energy markets, we treat them separately. Under an \ac{EACs} system, whether a miner has mined with green or brown energy is not directly related to miner's share of the grid mix or to the additional effect arising subsequently to the beginning of the miner's operations, but to whether the miner has purchased \ac{EACs} or not.

\subsection{Estimation methods for emissions per event}

Once the total emissions have been established, it is useful to identify an event that causally leads to the generation of these emissions, so as to establish the emissions per event, and to facilitate cost-benefit analysis based, for instance, on the event's benefits and the emissions' costs.

\paragraph{Origin accounting}

Origin accounting is the perspective that focuses on each ``coin", considering the environmental impact of each individual coin as given by the energy expenditures and energy mix used to mine them historically. In a way, it maintains an idea that the energy mix used to mine each given coin eternally accompanies the coin throughout its life, such that ``older" coins mined with little electricity will always be much greener than newer, more electricity-intensive coins. Under this paradigm, if tracing each coin to its origin is technically feasible, their economic fungibility is eroded.

However, holding bitcoin drives up bitcoin price, incentivizing additional mining. Whether the coin being held is old or new, green or not, has no bearing on this. In other words, even if it were possible to hold green bitcoin, doing so would create an incentive to mine more bitcoin, which may be green or brown. This incentive effect should be accounted for in a green Bitcoin framework.

\paragraph{Transaction accounting}

Transaction accounting is the practice of taking all the emissions (or electricity consumption) in a given time frame and dividing them by the number of transactions in the period, to arrive at a carbon (or electricity consumption) per transaction metric. This is usually framed to imply a causal effect from the transactions to the electricity consumption or emissions, and hence that additional transactions would entail additional emissions.

This method has received criticism because it overlooks the fact that, unlike in \ac{PoS} systems, electricity consumption from transaction processing is minimal in most \ac{PoW} systems. Instead, the overwhelming majority of the computational effort is derived from the mining process. Furthermore, by sidestepping Bitcoin's functioning as a wholesale payment system that can support a very large number of transactions through \ac{L2} solutions, as well as by failing to account for the Bitcoin halvings which regularly reduce incentives for mining, the predictions tying additional emissions with every additional transactions are, as a rule, seriously flawed
\cite{Paez2022RFITransition,Imran2018TheMining}.

\paragraph{Maintenance accounting}

To address the shortcomings outlined above, Cross and Bailey \cite{Cross2021GreeningOffsets} suggest the maintenance accounting approach, which the Crypto Carbon Ratings Institute denominates "holding-based approach" \cite{CCRI2022EnergyBlockchain}. This considers that all holding of bitcoin creates a proportional incentive to mine bitcoin for the period of the holding. One may additionally, and in a symmetric manner to the transaction-based approach, divide all the emissions over a time period by the percentage of holding of each holder during the time frame.\footnote{"All the carbon gets mapped to hodlers in proportion to the amount of Bitcoin they own." \cite{WBD2022CanCross}}

This method is gaining increasing praise as it reflects the economic reality of Bitcoin mining much more accurately. Nevertheless, it also encounters some limitations. The fact that transaction processing represents a negligible part of the \textit{computational effort} in the Bitcoin protocol does not mean that transaction processing is only responsible for equally negligible electricity consumption. This is because transaction processing itself may create an incentive to mine that is much more than proportional to the share of computational effort attributed to it.\footnote{Note also that although all non-lost bitcoin must be held at any given moment, including for transaction purposes, this is of little relevance from an incentive perspective (see next footnote).} Therefore, the maintenance accounting approach accurately attributes emissions caused by bitcoin \textit{holding} but fails to account for emissions caused by bitcoin \textit{transactions}.\footnote{Cross and Bailey \cite{Cross2021GreeningOffsets} argue that block rewards and fees are both denominated in bitcoin and thus depend on bitcoin's price, which in turn is influenced by the investors who hold bitcoin. While all of this is true, we do not expect transaction fees to be a function of bitcoin price simply because they are denominated in bitcoin. Rather, transaction fees are determined by the supply and demand of block space, which in turn will be given by the expected utility payers find in Bitcoin as a payment protocol. For practical purposes, it is better to think of block rewards as denominated in bitcoin, but of transaction fees as denominated in fiat.}

\paragraph{Hybrid accounting}
For a more integral approach, one may attribute holders emissions derived from the pursuit of the block subsidy (in proportion to their holdings, maintenance accounting) and emissions derived from the pursuit of transaction fees to transactions (transaction accounting). This hybrid approach has been proposed by the CCRI \cite{CCRI2022EnergyBlockchain} (but also Cross \cite{Cross2022FarTxs}):

\begin{quote}
``From an incentive perspective, miners receive both block subsidies and transaction fees. While transaction fees are paid by entities that execute transactions, it might not be intuitive why the block subsidy is paid by all holders. To understand the relationship between holders and block subsidies, we need to consider the creation of new currency. As miners propose new blocks, they are rewarded with new coins. While the supply of the currency inflates, the value of the overall currency stays the same. Therefore, the value of the individual coin is decreasing; the value of every holding in the respective cryptocurrency gets devalued; the difference in form of new coins is paid to the miner as the block subsidy. This results in an indirect payment of all holders towards the miners which in turn use this money to purchase electricity to run their mining devices. Therefore, all holders are responsible for the share of the block subsidy of the overall reward. The hybrid approach accounts for these phenomena and distributes the total network emissions to both all holders and all entities executing transactions. The share between holders and transactions is weighted by the respective share of both components." \cite[14]{CCRI2022EnergyBlockchain}
\end{quote}

\section{A survey of proposed solutions to Bitcoin emissions}
\label{sec:survey1}

Having covered frameworks to address Bitcoin \ac{GHG} emissions, we turn to frameworks to account for green bitcoin attributes.

\subsection{Colored coins}

A first solution to Bitcoin's energy "problem" is to distinguish between "green", "grey" and "brown" coins \cite{Green2022WillBitcoin,Robehmed2013SanitizingAccounts,Rothwell-Ferraris2021SharkBitcoin}. This amounts to storing a record of the energy type used to mine each particular coin on-chain.\footnote{Alternatives include HODLing only green "virgin" coins, see https://www.hiveblockchain.com/.} The proposal dates back to at least 2013, when the project "Coin Validation" was put forward \cite{Robehmed2013SanitizingAccounts} (and was received with strong criticism \cite{Bitcoinism2013IsCompanies,Back2013CoinBitcoin}). Similar proposals have been suggested for KYC/AML purposes \cite{delCastillo2018BitfuryBusiness}.

Colored coins suffer from two main problems. Firstly, distinguishing between green coins and grey/brown ones comes at the cost of substantially impairing bitcoin fungibility, which is a key element to the cryptocurrency's moneyness \cite{Cross2021GreeningOffsets,Back2013CoinBitcoin} (see also \cite{Kuhn2021TheBitcoin}). Secondly, usage of colored coins assumes "origin accounting" \cite{Cross2021GreeningOffsets}.

\subsection{Incentive offsets}

In 2021, Cross and Bailey \cite{Cross2021GreeningOffsets} formalized this perspective by suggesting an incentive offset system from a maintenance accounting perspective. Their framework suggests that, in order to completely offset any emissions caused by holding bitcoin, any bitcoin holder should co-invest in sustainable bitcoin mining (however the individual defines sustainable) in proportion to their holdings. Thus, if a person holds 1\% of all bitcoin, they should invest in sustainably mining 1\% of the total bitcoin hash rate.

The mechanism behind this proposal relates to the incentives given by holding and by mining bitcoin. First, holding bitcoin creates an incentive to mine bitcoin, as the price of bitcoin is driven upwards and mining profitability is increased. Second, mining bitcoin creates an incentive not to mine bitcoin, as the hash rate is driven upwards and mining profitability likewise decreases. Hence, Cross and Bailey suggest co-investing in green mining proportion to the holdings so as to neutralize any incentive effect given by the holding of bitcoin.

This proposal presents some strong advantages, including its adaptability to each person's own definition of sustainability, the lack of a need to know the global Bitcoin grid mix, and, most importantly, that this proposal is profitable as long as mining is also profitableas well. However, scaling this proposal may present some implementation challenges as it may collide with existing systems leading to double-counting of green attributes (see \ref{subsec:double}).

\subsection{Traditional environmental market instruments}
Insofar bitcoin mining leads to net positive \ac{GHG} emissions, a miner, holder or payer in bitcoin (depending on the estimation method per event used) may choose to green their bitcoin holdings by purchasing environmental instruments in the market.

\paragraph{Carbon offsets}
Carbon offsets are reductions of \ac{GHG} emissions or removal thereof. A carbon offset certificate is an instrument (a ``token" in the more general sense of the word) that not only certifies the offset but furthermore allows the bearer or owner to claim the offset as their own so as to compensate their emissions. A bitcoin miner with positive \ac{GHG} emissions may purchase carbon offsets to achieve, for instance, net carbon neutrality. Offset certificates may be considered environmentally beneficial in that the more they are demanded, the higher their price, the higher rewards for offsetors, and, thus, the higher the incentive to continue offsetting.

However, carbon offsets have received criticism:
\begin{enumerate}
    \item That they act as a ``free pass" for emitting parties to continue doing so and ``greenwash" their activities \cite{Harvey2023GreenwashingOffsetting,Raji2022IsGreenwashing,Gelmini2021WereGreenwash}.\footnote{A more charitable perspective may recognize the incentive generated by carbon offset purchases on additional offsetting projects but nevertheless find that offsets partially act as a free pass if low-quality offsets are cheaper than (and thus, prioritized over) highly reliable in-house \ac{GHG} reductions, in a Gresham's law of sorts \cite{Raji2022IsGreenwashing}.}
    \item That they are not trustworthy, and that the claimed quantities of ``sunk" \ac{GHG}s are grossly overestimated, would have been sunk irrespective of the offsetor's activity, that the emissions will ``bounce back" after a period, or that they suffer from other empirical problems such as double-counting, lack of additionally, etc. \cite{Anderson2012TheOffsets,Wara2008A74}
    \item That, from a financial perspective at least, they constitute a pure loss for the buyer, who gains no revenue from them \cite{Cross2021GreeningOffsets}.
\end{enumerate}

Note also that if a miner with \textit{negative} \ac{GHG} emissions wants to \textit{sell} carbon offsets to create a green premium on their hash rate and thus be incentivized to invest further in renewable mining, it may be unable to do so. This is because one of the main ways in which green miners may achieve net negative emissions is by either driving brown miners out of the market by increasing the global hash rate, or by facilitating additional renewable buildout through \ac{FLR} \cite{Ibanez2023CanSoK}. Both of these are second-order effects and subject to uncertainty and many other factors, and hence cannot be verified, audited and certified even if the incentive effect is there.\footnote{Note that this is not the case for methane miners, who can easily certify an emissions reduction at the only cost of assuming that methane is more ``harmful" than carbon dioxide (see \cite{Allen2018AMitigation,Cusworth2022WhenMetrics,Ibanez2023CanSoK}).}

\paragraph{Carbon credits}
Carbon credits are instruments similar to carbon offsets, except that they are usually the result of regulation and that they do not represent sunk a quantity of \ac{GHG} but rather a right to legitimately emit such a quantity. For this reason, a miner or holder may purchase them. They share some of the criticism received with carbon offsets, namely that they may enable greenwashing \cite{GurcamTheGreenwashing} and that they do not produce revenue for the buyer \cite{Cross2021GreeningOffsets}.\footnote{One could also argue that, unlike offset certificates, carbon credits do not produce in themselves any incentive for additional offsetting or renewable buildout as they do not directly reward offsetors or \ac{RES} producers. However, this is not obviously the case. A high price for carbon credits may incentivize firms to lower or offset emissions so as to avoid having to purchase the credits.}

\paragraph{\ac{RECs} and \ac{GOs}}
If purchasing energy from the grid, where green energy electrons and brown energy electrons are indistinguishable, a miner may purchase \ac{EACs} to be able to claim that its activities are green. This approach has the benefit that it is in line with how existing energy markets currently operate (i.e. through the commercialization of environmental attributes) and that it provides an incentive for renewable buildout as it effectively provides a "green premium" on electricity from \ac{RES}. \cite{Castellanos2017CryptocurrencyBlockchain}

However, \ac{EACs} have also been criticized under greenwashing pretenses. Specifically, it is argued that the incentive created by the green premium is not leading to additional renewable buildout (if there are new renewable investments these would be attributed to other factors, not income). This, the revenue from the sale of \ac{EACs} would be acting, effectively, as a windfall profit for \ac{RES} energy sellers whose facilities had already been built. However, an emissions reduction achieved directly by a company in the real world trades on par with an emissions reduction achieved indirectly through the purchase of \ac{EACs}, in spite of the latter leading to much less of the former. This would result, again, in a Gresham's Law analog of sorts, where bad emissions reductions displace good emissions reductions. \cite{Bjrn2022RenewableTargets,Brander2018CreativeEmissions,Gillenwater2014AdditionalityMarket,Gautam2021ProblematicGoals}

\subsection{Sui generis environmental Bitcoin market instruments}

We discussed above the difficulties miners face in selling the green attributes of their operations when these are net negative. To address this, proposals to tokenize these attributes have emerged. The Sustainable Bitcoin Protocol and Clean Incentive have designed Sustainable Bitcoin Certificates and Clean Bitcoin Certificates, respectively, each with some differentiating factors. These tokens store on-chain (on the Bitcoin-based Stacks blockchain and on ordinals-based Bitcoin inscriptions, respectively) a record that a coin has been mined sustainably, producing one token for every coin sustainably mined (following a verification process). A portfolio manager may green their bitcoin portfolio by holding one token for every coin held.\footnote{These systems includes tokenomic schemes to green not just new coins but also coins already mined.}\footnote{See \url{https://www.sustainablebtc.org/} and \url{https://www.cleanincentive.com/}.}

The benefits of these instruments are that they offer an additional revenue stream for the miner, that they are consistent with existing systems (especially the markets for environmental attributes) but also tailored to the specificities of bitcoin (these tokens may be minted for methane mining, off-grid mining, \ac{EACs}-based mining, etc.), and that they facilitate \ac{ESG} reporting. In turn, one of the challenges faced by these projects is the construction of technically sound ecosystems with sufficient buyers to make these tokens liquid.

\section{Discussion}

Throughout this paper, we have surveyed and systematized the various elements composing the systems to green Bitcoin and to account therefor. We have highlighted the pros and cons of different approaches. In the remainder, we interpret these findings and extract our key takeaways.

\subsection{Second-order causal effects and incentives}

Our research shows that properly accounting for the degree of sustainability of Bitcoin requires looking beyond the immediate, proximate short-term effects of a given event. Not doing so obscures important facts, such as how marginal electricity consumption (even if high in marginal emissions), how flexible loads, how carbon offsets and even how carbon credits may incentivize renewable buildout.  Similarly, it could be the case that pessimism about the usefulness of carbon credits, carbon offsets and \ac{EACs} is the result of a short-term bias and a neglect of the long-term incentives at play.

\subsection{Philosophical presuppositions}
An additional finding is the often-overlooked relevance of philosophical assumptions in the analysis of Bitcoin's "greenness" or ``brownness." If applied to attribute responsibility (instead of other, much more narrowly defined applications of the method), marginal emissions accounting is tied to the philosophical position that the legitimacy of an energy buyer is given by its relative age, a rather heterodox stance. This could suggest that an attributional framework is more adequate for this goal, whether through the egalitarianism of a simple average or the coherence of the \ac{EACs}-based methods. Overall, it is of paramount importance to apply a philosophical perspective consistently, both in all the dimensions of the analysis of a single industry as well as across industries.

\subsection{Coherence with the existing system}
The initiatives identified in this paper highlight the importance, for any Bitcoin greening scheme, of coherence with existing systems already in place, such as the market for environmental attributes. Frameworks developed that do not take them into account risk implementation obstacles and slower adoption.

\subsection{The problem of double-counting}
\label{subsec:double}
These constitute a genre of the former. The overlap of multiple frameworks can generate an ironic consequence for the technology created to solve the double-spending problem: the double-spending of environmental attributes. If a miner claims that their activities are green due to the energy mix of the grid to which they have plugged, but simultaneously \ac{EACs} have been sold for all the energy in the grid, there is double-counting as the greenness of the energy mix has been spent twice, once in the average attributional scheme and once in the \ac{EACs} scheme. Similarly, if an individual seeks to neutralize their bitcoin holding with Cross's incentive offset scheme by investing in a miner that claims the grid mix but did not purchase \ac{RECs}, the greenness of green energy may also be spent twice.\footnote{This circles back to the original issue of philosophical presuppositions: the attribution of responsibility between the two parties spending an attribute twice is not entirely obvious.}

\subsection{The need for an \ac{ESG} reporting framework}
Carbon-negative second-order effects in Bitcoin must be accounted for in manners that adequately fit existing reporting practices (as per the previous discussion of ``coherence"). A potentially fruitful avenue to this end would be the application of the concept of ``Scope 4" emissions (avoided emissions) \cite{CDSB2020ScopeAction,PwC2022ScopeImpact}.

We previously established that second-order avoided emissions cannot be reliably ``verified", but that the incentive effect to avoid is real nonetheless. The additional renewable buildout that is incentivized through existing instruments such as \ac{RECs} or carbon offsets also cannot be demonstrated, and yet we admit e.g. \ac{EACs} as a valuable tool in carbon accounting. A similar understanding could guide Scope 4-based instruments.

\section{Conclusion}

We surveyed and systematized the state of the art pertaining to Bitcoin carbon accounting. As a result, we are inclined to highlight the importance of making philosophical assumptions clear, considering second-order effects, integrating any green Bitcoin frameworks into existing schemes, and avoiding double-counting.

\section*{Acknowledgements}
\label{sec:acknowledgements}

We thank Troy Cross, Casey Martinez, and Elliot David for comments that greatly improved the manuscript.
J.I.I. was supported by the DLT Science Foundation and the University College London Centre for Blockchain Technologies. A.F. was supported by the University College London Centre for Blockchain Technologies and Energiequelle GmbH.

%\section*{Author Contributions}
%...

\section*{Conflict of Interest}

The authors declare that they have no known competing financial interests or personal relationships that could have appeared to influence the work reported in this paper.

\printacronyms

\printbibliography

%\newpage

\raggedbottom

\vspace{1em}

%\appendix

%\section*{Appendix...}
%Pending

\end{document}